\documentclass[11pt]{iopart}

\usepackage{bm}
\usepackage{srcltx} 
\usepackage{color}

\usepackage{setstack,amssymb,amsfonts,graphicx,iopams}

\begin{document}

\title{
Finite-temperature effective boundary theory of the quantized thermal Hall effect
}
\date{\today}
\author{Ryota Nakai$^1$, Shinsei Ryu$^2$, Kentaro Nomura$^3$}
\address{$^1$ WPI-Advanced Institute for Materials Research (WPI-AIMR), Tohoku University,
Sendai 980-8577, Japan}
\address{$^2$ Department of Physics, University of Illinois, 
1110 West Green St, Urbana IL 61801}
\address{$^3$ Institute for Materials Research,
Tohoku University, Sendai 980-8577, Japan}
\ead{rnakai@wpi-aimr.tohoku.ac.jp}

\begin{abstract}
A finite-temperature effective free energy of the boundary of a quantized thermal Hall system is derived microscopically from the bulk two-dimensional Dirac fermion coupled with a gravitational field.
In two spatial dimensions, 
the thermal Hall conductivity of fully gapped insulators and superconductors is quantized 
and given by the bulk Chern number, in analogy to the quantized electric Hall conductivity in quantum Hall systems.
From the perspective of effective action functionals, two distinct types of the field theory have been proposed to describe the quantized thermal Hall effect. One of these,
known as the gravitational Chern-Simons action, is a kind of topological field theory,
and the other is a phenomenological theory
relevant to the Str\v{e}da formula.
In order to solve this problem, we derive microscopically an effective theory that accounts for the quantized thermal Hall effect.
In this paper, the two-dimensional Dirac fermion under a static background gravitational field is considered in equilibrium at a finite temperature, from which an effective boundary free energy functional of the gravitational field is derived.
This boundary theory is shown to explain the quantized thermal Hall conductivity and thermal Hall current in the bulk by assuming the Lorentz symmetry.
The bulk effective theory is consistently determined via the boundary effective theory.
\end{abstract}

\submitto{\NJP}

\maketitle

\section{Introduction}

Topology of the energy band structure in insulators and superconductors emerges in transport phenomena,
and the topological number can be detected as a transport coefficient\cite{hasan10,qi11}.
The electric Hall conductivity is quantized and given by the Chern number of the occupied wave functions in two-dimensional band insulators\cite{prange87,thouless82}. 
The Chern number is nonzero when an electronic system is in time-reversal symmetry broken topological phases.
Similarly, two-dimensional insulators and superconductors show the quantized thermal Hall effect in time-reversal symmetry broken 
topological phases\cite{read00}.
This similarity between electric and thermal responses reflects 
a parallel between topological insulators in symmetry class A and topological superconductors in symmetry class D, 
both of which are characterized, in two dimensions, by the Chern number of the filled energy bands\cite{schnyder08,kitaev09,ryu10}.

Topological quantum field theories are efficient descriptions and characterizations of topological phases.
Effective actions for the external electromagnetic and gravitational fields
are given in terms of topological terms and attributed to quantum anomalies
\cite{bertlmann96,fujikawa04}. 
They can be used to discuss the classification of topological insulators and superconductors in arbitrary dimensions\cite{qi08,wang11,ryu12}.
Quantized electromagnetic responses, including the quantum Hall effect in two dimensions, 
can be viewed as responses resulting from the Chern-Simons action functional. 
For the case of the electromagnetic field, the Chern-Simons action is given as
\begin{eqnarray}
 S^{\rm{EM}}[A]
 =
 \frac{Ce^2}{4\pi}
 \int \rmd^3x\,
 \epsilon^{\mu\nu\rho}
 A_{\mu}\partial_{\nu}A_{\rho}
 \label{eq:electromagneticCS}
\end{eqnarray}
in the (2+1)-dimensional space-time, where $C$ is the Chern number (here
and henceforth natural units with $c=\hbar=k_B=1$ are used).
On the other hand, 
the quantized thermal Hall effect has been predicted to occur in a weak pairing phase of the chiral p-wave superconductor in Ref.~\cite{read00}, 
which is considered to be a realization of a two-dimensional time-reversal symmetry broken topological superconductor. 
In analogy with the electromagnetic response, it has been claimed that the quantized thermal response of topological superconductors is described by the gravitational Chern-Simons 
action\cite{read00}
\begin{eqnarray}
 S^{\rm{G}}[\omega]
 &=
 \frac{C}{96\pi}
 \int \rmd^3x\,
 \epsilon^{\mu\nu\rho} 
 \tr
 \left(
 {\omega_{\mu}}
 \partial_{\nu}
 {\omega_{\rho}}
 +
 \frac{2}{3}
 {\omega_{\mu}}
 {\omega_{\nu}}
 {\omega_{\rho}}
 \right),
 \label{eq:gravitationalCS}
\end{eqnarray}
where $\omega_{\mu}$ is the spin connection.
In recent years,
space-time curvature is widely used to study characteristic responses of topological materials\cite{hughes11,gromov15,bradlyn15}. 
The gravitational Chern-Simons action is associated with the gravitational anomaly in (2+1) dimensions
\cite{bertlmann96,fujikawa04,alvarez-gaume84}, and can be microscopically derived from the (2+1)-dimensional 
massive Dirac fermion coupled with the background
gravitational field at zero temperature\cite{vuorio86,goni86,ojima89}.
However, it has been noted
that a thermal Hall current in the bulk cannot be created by a uniform gravitational field gradient as a response derived from the gravitational Chern-Simons action\cite{stone12}.
Therefore, while the coefficient of the gravitational Chern-Simons term (``the chiral central charge'') 
is related to the quantized thermal Hall conductance, 
the quantized thermal transport and the gravitational Chern-Simons term appear to be more remotely related
then the quantized electromagnetic response and the $U(1)$ Chern-Simons term.

An inhomogeneous temperature field driving thermal transport is effectively realized 
by a gravitational potential field $\phi$ 
through the Luttinger's phenomenological argument using the Tolman-Ehrenfest relation\cite{luttinger64}
\begin{eqnarray}
 \frac{1}{T}\bm{\partial}T
 =
 \bm{\partial}\phi.
 \label{eq:Luttinger_temperature_gravity}
\end{eqnarray}
Owing to the relation $j_T=j_E-(\mu/e)j_C$, the thermal current $j_T$ is identified as
the energy current $j_E$ when the charge current $j_C$ does not contribute at zero chemical potential $\mu=0$, which is true for superconductors.
A thermal current induced by a temperature gradient is then equivalent to an energy current induced by the space-time metric $g_{\mu\nu}$.

From the phenomenological point of view, an analogy between the electric and the thermal transport holds. 
The Wiedemann-Franz law connecting the electric and the thermal Hall conductivity for the Dirac fermion\cite{qin11,nomura12,sumiyoshi13}  has been proved to be
\begin{eqnarray}
 \kappa_{H}
 =
 \frac{\pi^2}{3}\frac{T}{e^2}\sigma_H.
 \label{eq:wiedemann_franz}
\end{eqnarray}
Note that in the case of topological superconductors, the coefficient is half of that in (\ref{eq:wiedemann_franz}) since a Majorana fermion, 
a quasi-particle in a topological superconductor, is half of a complex fermion. 
Based on the phenomenological analogy of the Hall conductivity, an effective free energy for the quantized thermal Hall effect
\begin{eqnarray}
 j_T^k=\kappa_H\epsilon^{kl}\partial_l T
 \label{eq:thermal_Hall_effect}
\end{eqnarray}
 of a Lorentz invariant system has been proposed as\cite{nomura12}
\begin{eqnarray}
 F[\phi,\Omega]
 \propto
 \frac{\kappa_{H}T}{v^2}
 \int \rmd^2x
 \phi\Omega,
 \label{eq:effective_phenomenology}
\end{eqnarray}
where $\Omega$ is the angular velocity of the system in a rotating frame, and $\kappa_H=C\pi^2T/6\pi$ is the thermal Hall conductivity for insulators whose occupied energy bands have a total Chern number $C$, and $v$ is the Fermi velocity of the Dirac fermion.
This form of the phenomenological effective free energy (\ref{eq:effective_phenomenology}) can be realized when we consider a metric in a rotating frame with a gravitational potential.
So far, a action or a free energy of the gravitational field that accounts for the quantized thermal Hall effect has not been derived directly from fermionic models of topological insulators and topological superconductors.

In this study, an effective free energy functional of the gravitational field is derived microscopically from the two-dimensional massive Dirac fermion coupled with the static gravitational field.
The effective free energy is derived by the following procedure.
First, we consider the gapless boundary fermion, which is a manifestation of the Dirac fermion 
with nontrivial bulk energy band topology.
Then an effective free energy of the boundary theory is calculated by the field theoretical method. 
We also consider the bulk thermal Hall current and a bulk effective field theory.

This paper is organized as follows.
As a preliminary to the main contents of this study, phenomenological theory of the quantized thermal Hall effect is reviewed in Sec.~\ref{sec:phenomenology}, and a microscopic model studied in this paper is introduced in Sec.~\ref{sec:model}.
In Sec.~\ref{sec:projection} the two-dimensional bulk fermion Hamiltonian is projected onto the one-dimensional boundary to give the boundary fermion Hamiltonian, and, in Sec.~\ref{sec:boundary_FE}, the boundary effective free energy is calculated from the boundary fermion. 
In Sec.~\ref{sec:bulk_conductivity}, the quantized thermal Hall conductivity is derived from the boundary theory.
In Sec.~\ref{sec:bulk_current}, the thermal Hall current of the two-dimensional gapped bulk are deduced from the boundary effective free energy.
In Sec.~\ref{sec:bulk_FE}, the bulk effective free energy is conjectured from the expression of the bulk thermal current.
We give discussion about our results in Sec.~\ref{sec:discussion} and summarize our conclusions in Sec.~\ref{sec:conclusion}.

\section{Preliminaries}
\label{sec:model_method}

\subsection{Phenomenology}
\label{sec:phenomenology}

In this subsection, we briefly review the phenomenology of the thermal Hall effect\cite{qin11,nomura12} of a Lorentz invariant system in two dimensions.
Quantized thermal (energetic) Hall current in a two-dimensional fully gapped topological insulator and superconductor is phenomenologically defined in analogy with the quantized part of the electric Hall current in the quantum Hall system.
The electric Hall current is defined by
\begin{eqnarray}
 \bm{j}_C
 =
 \sigma_H
 \hat{\rm{z}}
 \times
 \bm{E},
 \label{eq:Hall_current}
\end{eqnarray}
where $\hat{\rm{z}}$ is the unit vector pointing the out-of-plane direction.
Combining (\ref{eq:Hall_current}) with the Str\v{e}da formula\cite{streda82} for the quantized electric Hall conductivity, given by the derivative of the magnetization with respect to the chemical potential
\begin{eqnarray}
 \sigma_H
 =
 -e
 \frac{\partial M^z}{\partial\mu},
\end{eqnarray}
and a phenomenological relation $e\bm{E}=-\bm{\partial}\mu$, we have obtained the relation
\begin{eqnarray}
 \bm{j}_C
 =
 \bm{\partial}\mu
 \times
 \frac{\partial \bm{M}}{\partial\mu}.
 \label{eq:magnetization_current}
\end{eqnarray}
The electric current (\ref{eq:magnetization_current}) can be regarded as a magnetization current $\bm{j}_C=\bm{\partial}\times\bm{M}$, if we assume that the magnetization varies according to variation of the chemical potential.
Similarly, 
by using the Str\v{e}da formula for the quantized thermal Hall conductivity\cite{nomura12}, given by the derivative of the energy magnetization $\bm{M}_E$ with respect to temperature
\begin{eqnarray}
 \kappa_H
 =
 -\frac{\partial M^z_E}{\partial T},
 \label{eq:Streda_thermal_Hall}
\end{eqnarray}
the thermal Hall current 
\begin{eqnarray}
 \bm{j}_E
 =
 -\kappa_H
 \hat{\rm{z}}
 \times
 \bm{\partial}T
 \label{eq:thermal_Hall_current}
\end{eqnarray} 
can be identified with the energy version of the magnetization current\cite{qin11} $\bm{j}_E=\bm{\partial}\times\bm{M}_E$, when the energy magnetization changes through variation of temperature as
\begin{eqnarray}
 \bm{j}_E
 =
 \bm{\partial}T
 \times
 \frac{\partial \bm{M}_E}{\partial T}.
 \label{eq:energy_current_energy_magnetization}
\end{eqnarray}
Note that $\bm{M}$ and $\bm{M}_E$ have only $z$-component.
This fact implies that the thermal Hall current in a fully gapped system can be described in terms of the magnetization type energy current caused by the energy magnetization.

Lorentz invariance of the fermionic system gives rise to a symmetry of the energy-momentum tensor with respect to permutations of the indices.
The energy momentum tensor has the form of
\begin{eqnarray}
 T^{\mu\nu}
 =
\left (
 \begin{array}{ccc}
  \epsilon & j_E^{1}/v & j_E^{2}/v  \\ 
  v\pi^1 & \Sigma^{11} & \Sigma^{12}  \\
  v\pi^2 & \Sigma^{21} & \Sigma^{22}  
 \end{array}
 \right),
 \label{eq:energy_momentum_tensor}
\end{eqnarray}
where $\epsilon$ is the energy density, $\pi^j$ is the momentum density, and $\Sigma^{jk}$ is the stress tensor. 
The symmetry of the tensor gives the equation $\bm{j}_E=v^2\bm{\pi}$, and then it leads to an equality between the energy magnetization $\bm{M}_E^{\mu\nu}=\langle x^{\mu}j_E^{\nu}-x^{\nu}j_E^{\mu}\rangle/2$ and the angular momentum $\bm{L}^{\mu\nu}=\langle x^{\mu}\pi^{\nu}-x^{\nu}\pi^{\mu}\rangle$ as
\begin{eqnarray}
 M_E^{\mu\nu}=v^2L^{\mu\nu}/2,
 \label{eq:energy_magnetization_angular_momentum}
\end{eqnarray} 
where $M_E^z=M_E^{12}$ and $L^z=L^{12}$.
Here $\langle\cdots\rangle$ indicates evaluating an operator in thermal equilibrium at temperature $T$.
Substituting the equation (\ref{eq:energy_current_energy_magnetization}) and (\ref{eq:energy_magnetization_angular_momentum}) into the definition of the thermal Hall coefficient, we obtain
\begin{eqnarray}
 \kappa_{H}
 =
 -\frac{j_E^x}{\partial_y T}
 =
 -\frac{v^2}{2}
 \frac{\partial L^{z}}{\partial T}
 = 
 -\frac{v^2}{2T}
 \frac{\partial L^{z}}{\partial \phi},
 \label{eq:th_Hall_angular_momentum}
\end{eqnarray}
where in the last equality of (\ref{eq:th_Hall_angular_momentum}), the relation (\ref{eq:Luttinger_temperature_gravity}) is used.
The angular momentum is the conjugate field of the angular velocity $\Omega$, and is given by variation of the free energy as $L^{z}=-\delta F/\delta\Omega^z$.
The free energy functional for the thermal Hall effect is given by
\begin{eqnarray}
 F[\phi,\Omega]
 =
 -\int \rmd^2x
 L^{z}\Omega
 =
 \int \rmd^2x
 \frac{2\kappa_H T}{v^2}
 \phi\Omega.
\end{eqnarray}
Corresponding electromagnetic free energy is the electromagnetic Chern-Simons term with gauge fixing $F^{\rm{EM}}\propto\sigma_H\int \rmd^2x A_0B^z$. 
The free energy $F^{\rm{EM}}$ 
represents an quantized electric Hall response, that is, 
an electric current is induced by a magnetization given by $M^z=-\delta F^{\rm{EM}}/\delta B^z$.

\subsection{Two-dimensional Dirac fermion under gravitational field}
\label{sec:model}

The Dirac fermion with the Fermi velocity $v$ is invariant under the Lorentz transformation which preserves the line element
\begin{eqnarray}
 \rmd s^2=(v\rmd t)^2-\rmd x^2-\rmd y^2.
\end{eqnarray}
Introducing the three coordinates $x^{\mu}=(vt,x,y)$, 
a metric representing a system under a gravitational potential $\phi$ rotating with an angular velocity $\Omega$ is given by\cite{landau75}
\begin{eqnarray}
 g_{\mu\nu}
 =
 \left(
 \begin{array}{ccc}
  1+2\phi & \Omega y/v & -\Omega x/v \\
  \Omega y/v & -1 & 0 \\
  -\Omega x/v & 0 & -1
 \end{array}
 \right),
 \label{eq:metric_potential_rotation}
\end{eqnarray}
up to linear order in $\phi$ and $\Omega$.
The effective field theory addressed in this paper is the free energy functional of the metric that results from the two-dimensional Dirac fermion coupled with the gravitational field.
In the following, we proceed with a general metric form, and 
the metric (\ref{eq:metric_potential_rotation}) is substituted into the expression when the Tolman-Ehrenfest relation (\ref{eq:Luttinger_temperature_gravity}) is used to mimic temperature gradient.

The two-dimensional Dirac fermion Hamiltonian coupled with the gravitational field 
is separated into the flat and the remaining parts as follows. 
Deviation of a curved space-time metric from the flat Minkowski space-time is defined by
\begin{eqnarray}
 g_{\mu\nu}
 =
 \eta_{\mu\nu}
 +
 h_{\mu\nu}.
\end{eqnarray}
We use the sign convention $\eta_{\mu\nu}={\rm diag}(+1,-1,-1)$.
The metrics $g_{\mu\nu}$ and $\eta_{\mu\nu}$ are related by a triad field $e_{\;\;\mu}^{\alpha}$ with the equation 
\begin{eqnarray}
 g_{\mu\nu}
 =
 e_{\;\;\mu}^{\alpha}
 e_{\;\;\nu}^{\beta}
 \eta_{\alpha\beta}.
\end{eqnarray} 
When a deviation is small enough ($h_{\mu\nu}\ll 1$), 
we can write the triad field in terms of $h$ as $e_{\;\;\mu}^{\alpha}\simeq \delta_{\;\;\mu}^{\alpha}+h_{\;\;\mu}^{\alpha}/2$.
Subscripts and superscripts of $h_{\mu\nu}$ are raised and lowered by $\eta_{\mu\nu}$ and its inverse $\eta^{\mu\nu}$, like $h_{\;\;\mu}^{\alpha}=h_{\nu\mu}\eta^{\alpha\nu}$.
The inverse and the determinant of the triad field are given, up to linear order in $h$, by
\begin{eqnarray}
 &e^{\;\;\mu}_{\alpha}
 \simeq
 \delta^{\mu}_{\alpha}
 -
 h^{\;\;\mu}_{\alpha}/2,\\
 &\sqrt{g}
 =
 \sqrt{{\rm det}\,g_{\mu\nu}}
 =
 {\rm det}(e_{\;\;\mu}^{\alpha})
 \simeq
 1+h/2,
\end{eqnarray}
where $h=h^{\mu}_{\;\;\mu}$. 
The inverse of the triad satisfies 
$e_{\;\;\mu}^{\alpha}e^{\;\;\nu}_{\alpha}=\delta_{\mu}^{\nu}$,
$e_{\;\;\mu}^{\alpha}e^{\;\;\mu}_{\beta}=\delta^{\alpha}_{\beta}$, 
and $e^{\;\;\mu}_{\alpha}e^{\;\;\nu}_{\beta}g_{\mu\nu}=\eta_{\alpha\beta}$.
The gamma matrices on the curved space-time $\underline{\gamma}^{\mu}$ satisfy the anticommutation relation $\{\underline{\gamma}^{\mu},\underline{\gamma}^{\nu}\}=2g^{\mu\nu}$. 
$\underline{\gamma}^{\mu}$ is related to the gamma matrices on the flat space-time $\gamma^{\alpha}$ by 
\begin{eqnarray}
 \underline{\gamma}^{\mu}=e_{\alpha}^{\;\;\mu}\gamma^{\alpha},
\end{eqnarray}
where $\gamma^{\alpha}$ satisfies the relation $\{\gamma^{\alpha},\gamma^{\beta}\}=\eta^{\alpha\beta}$.

The action of the (2+1)-dimensional Dirac fermion coupled with a gravitational field is 
\begin{eqnarray}
 S
 =
 \int \rmd t\rmd^2x
 \sqrt{g}\,
 \bar{\psi}
 \left[
 \rmi v\underline{\gamma}^{\mu}
 \nabla_{\mu}
 -
 m
 \right]
 \psi.
\end{eqnarray}
Here the covariant derivative $\nabla_{\mu}$ of the spinor field is given by $\nabla_{\mu}\psi=(\partial_{\mu}-\omega_{\mu})\psi$,
where the connection is given by 
$\omega_{\mu}=-\gamma_{\alpha\beta}\omega_{\mu}^{\;\;\alpha\beta}/4$, 
using 
$\gamma_{\alpha\beta}=[\gamma_{\alpha},\gamma_{\beta}]/2$ 
with 
$\gamma_{\alpha}=\eta_{\alpha\beta}\gamma^{\beta}$,
and 
\begin{eqnarray}
 \omega_{\mu}^{\;\;\alpha\beta}
 =
 e^{\beta}_{\;\;\nu}g^{\nu\lambda}(\partial_{\mu}e^{\alpha}_{\;\;\lambda}
 -
 \Gamma^{\rho}_{\;\;\mu\lambda}e^{\alpha}_{\;\;\rho}).
 \label{eq:spin_connection_component}
\end{eqnarray}
The local Hamiltonian $\mathcal{H}$ is defined by
\begin{eqnarray}
 S
 =
 \int \rmd t\rmd^2x
 \sqrt{g}\,
 \psi^{\dagger}
 \gamma^0
 \underline{\gamma}^{0}
 \left(
 \rmi\partial_t
 -
 \mathcal{H}
 \right)
 \psi.
\end{eqnarray}
Then the
Hamiltonian of the two-dimensional Dirac fermion coupled with a gravitational field is written as
\begin{eqnarray}
 H
 &=
 \int \rmd^2x\,
 \psi^{\dagger}
 \sqrt{g}
 \gamma^0
 \underline{\gamma}^{0}
 \mathcal{H}
 \psi \nonumber\\
 &=
 \int \rmd^2x\,
 \psi^{\dagger}
 \sqrt{g}
 \left[
 \rmi v\gamma^0\underline{\gamma}^0\omega_0
 -
 \rmi v\gamma^0
 \underline{\gamma}^j
 \nabla_j
 +
 m\gamma^0
 \right]
 \psi 
 \label{eq:Hamiltonian_explicit}\\
 &\simeq
 \int \rmd^2x\,
 \psi^{\dagger}
 [\mathcal{H}_0
 +
 \mathcal{U}]
 \psi.
 \label{eq:Hamiltonian}
\end{eqnarray}
Latin indices in (\ref{eq:Hamiltonian_explicit}) run over the spatial dimensions ($j=1,2$) and Greek indices run over both temporal and spatial dimensions ($\alpha=0,1,2$).
Matrix element of the unperturbed Hamiltonian
is given by
\begin{eqnarray}
 &\mathcal{H}_0
 =
 -\rmi v
 \gamma^0\gamma^j
 \partial_j
 +
 m\gamma^0.
 \label{eq:Hamiltonian_unperturbed}
\end{eqnarray}
The perturbation Hamiltonian consists of two parts $\mathcal{U}=\mathcal{H}_1+\mathcal{H}_2$; one from variation of the triad and metric in (\ref{eq:Hamiltonian_explicit}), and the other from the spin connection. The matrix elements of the perturbation terms are, respectively,
\begin{eqnarray}
 &\mathcal{H}_1
 =
 (h/2)
 \mathcal{H}_0
 +\rmi v
 (h^{\;\;j}_{\alpha}/2)
 \gamma^0\gamma^{\alpha}
 \partial_j,
 \label{eq:Hamiltonian_perturbation}\\
 &
 \mathcal{H}_2
 =
 -(\rmi v/4)\gamma^0
 \gamma^{\mu}
 \gamma^{\alpha}\gamma^{\beta}\cdot
 \omega^{(1)}_{\mu\;\alpha\beta},
 \label{eq:Hamiltonian_perturbation2}
\end{eqnarray}
where $\omega^{(1)}_{\mu\;\alpha\beta}$ is a part of $\omega_{\mu\;\alpha\beta}=\eta_{\alpha\gamma}\eta_{\beta\delta}\omega_{\mu}^{\;\;\gamma\delta}$ that contains at most linear order in $h^{\alpha}_{\;\;\nu}$, and we have used $\omega_{\mu\;\alpha\beta}=-\omega_{\mu\;\beta\alpha}$.
Later the second perturbation term (\ref{eq:Hamiltonian_perturbation2}) will be dropped out since it does not contribute to the finite temperature free energy.
Although the original Lagrangian has the Lorentz invariance, symmetries of the Hamiltonian (\ref{eq:Hamiltonian_unperturbed}) and (\ref{eq:Hamiltonian_perturbation}) are reduced down to the SO(2) rotational invariance in the $x^1$-$x^2$ plane. 
Time-reversal symmetry and parity symmetry are broken by the mass term.

\section{Effective theory at the boundary}
\label{sec:boundary_theory}

In the (2+1)-dimensional space-time, the Chern-Simons action is induced by the parity breaking mass term of the Dirac fermion.
The Chern-Simons action of the U(1) gauge field is derived as an effective action by tracing out the Dirac fermionic degrees of freedom of the action of the (2+1)-dimensional Dirac fermion coupled with an electromagnetic field\cite{niemi83,redlich84l,redlich84d}.
Similarly, the gravitational Chern-Simons action, the Chern-Simons action of the spin connection, is derived by tracing out the fermionic degrees of freedom of the action of the Dirac fermion coupled with the gravitational field\cite{vuorio86,goni86,ojima89}.

When the theory of induced Chern-Simons terms is extended to a finite temperature system by considering the imaginary time, the induced theory gives the partition function and the free energy of the gauge fields.
However, if the fermionic system is fully gapped, temperature dependent part of the free energy is exponentially suppressed by the factor $\rme^{-\beta|m|}$, where $\beta$ is the inverse temperature and $|m|$ is the magnitude of the gap\cite{babu87,manes13}.
The effective free energy of the quantized thermal Hall effect will not be given in this way, since the quantized thermal Hall conductivity is linearly dependent on temperature.
Therefore in order to derive the finite-temperature effective free energy for the quantized thermal Hall effect, we, at first, consider the gapless boundary fermion, the existence of which is guaranteed by the nontrivial energy band topology of gapped fermionic systems.
The effective free energy of the gravitational field for the one-dimensional boundary modes is calculated by the field theoretical method.
This calculation procedure seems consistent with the fact that the boundary theory is legitimate to study the quantized thermal Hall effect\cite{kane97}.

For the purpose of deriving a finite temperature effective free energy, we consider the finite temperature path integral. 
Tracing out fermionic degrees of freedom of the density matrix gives an effective free energy functional of the gravitational field as
\begin{eqnarray}
 \exp(-\beta F[h])
 &=
 \int 
 \mathcal{D}\psi^{\dagger}
 \mathcal{D}\psi
 \exp(-\beta H[\psi^{\dagger},\psi,h]) \nonumber\\
 &=
 \prod_n{\rm Det}\,\mathcal{G}^{-1}(\rmi\omega_n),
\end{eqnarray}
where $\mathcal{G}(\rmi\omega_n)=1/(-\rmi\omega_n+\mathcal{H}_0+\mathcal{U})$ is the temperature Green's function with the Matsubara frequency 
for fermions $\omega_n=(n+1/2)2\pi/\beta\,(n\in\mathbb{Z})$, and $\rm{Det}$ is taken for the Hilbert space on the two-dimensional space and $2\times 2$ matrix degrees of freedom.
Expanding the temperature Green's function with respect to $\mathcal{U}$, we obtain 
\begin{eqnarray}
 \ln\mathcal{G}^{-1}
 &=
 \ln\mathcal{G}_0^{-1} 
 +
 \ln(1+\mathcal{G}_0\mathcal{U}) \nonumber\\
 &=
 \ln\mathcal{G}_0^{-1} 
 +
 \sum_{l=1}^{\infty}
 \frac{1}{l}(\mathcal{G}_0\mathcal{U})^l,
\end{eqnarray}
where $\mathcal{G}_0(\rmi\omega_n)=1/(-\rmi\omega_n+\mathcal{H}_0)$ is the temperature Green's function in the flat space-time. 
Then expansion of the effective free energy with respect to $h_{\mu\nu}$, that is $F=\sum_{l=0}^{\infty}F^{(l)}$, is given by
\begin{eqnarray}
 &-\beta
 F^{(0)}
 =
 \sum_n \Tr
 \ln\mathcal{G}_0^{-1}, \\
 &-\beta
 F^{(l)}
 =
 \frac{1}{l}
 \sum_n
 \Tr
 \left(\mathcal{G}_0\mathcal{U}\right)^l,
 \label{eq:effectiveFE_perturbation_2}
\end{eqnarray} 
where the identity $\rm{ln}\,\rm{Det}=\rm{Tr}\,\rm{ln}$ is used, and $\rm{Tr}$ is taken for the same Hilbert space as $\rm{Det}$.
$F^{(0)}$ is independent of the gravitational field and gives only a constant, and the second order perturbation term $F^{(2)}$ is the focus of this paper.

\subsection{Boundary fermion}
\label{sec:projection}

In two-dimensional space, consider a boundary at $x^1=0$ between a gapped bulk at $x^1<0$ with mass $m$ and that at $x^1>0$ with mass $-m$.
The boundary is extended to an entire $x^2$ space.
The masses in the two gapped semi-infinite regions
are smoothly connected by introducing an $x^1$-dependent mass term, 
whose sign changes at the boundary (${\rm sgn}[m(x^1<0)]=-{\rm sgn}[m(x^1>0)]$).
Since the Chern number at both side of the boundary differs by unity, the boundary hosts a single chiral boundary mode of the massless Dirac fermion.
The unperturbed Hamiltonian (\ref{eq:Hamiltonian_unperturbed}) is decoupled into 
the $x^1$- and the $x^2$-dependent parts.
The wave function of the boundary mode of the Hamiltonian (\ref{eq:Hamiltonian_unperturbed}) is a product of a plane wave of the $x^2$-coordinate and a two-components spinor wave function of the $x^1$-coordinate satisfying
\begin{eqnarray}
 \left(
 -\rmi v
 \gamma^0\gamma^1
 \partial_1
 +
 m(x^1)\gamma^0
 \right)
 \psi_0(x^1)
 =0.
 \label{eq:boundary_eq}
\end{eqnarray}
Formally the solution of (\ref{eq:boundary_eq}) is given by
\begin{eqnarray}
 \psi_0(x^1)
 =
 \langle x^1|\psi_0\rangle
 =
 \frac{1}{\sqrt{\mathcal{N}}}
 \exp
 \left[\rmi\gamma^1v^{-1}
 \int_0^{x^1} \rmd{x'}^1\, m({x'}^1)
 \right]
 |s\rangle,
 \label{eq:boundary_state_wavefunction}
\end{eqnarray}
where we have used an identity $(\gamma^1)^2=\eta^{11}=-1$,
and the normalization constant is defined by $\mathcal{N}=\int dx^1 \psi_0^{\dagger}(x^1)\psi_0(x^1)=\langle\psi_0|\psi_0\rangle$.
Here, $|x^1\rangle$ is an eigenstate vector of the position operator $\hat{x}^1$ in the Hilbert space on the $x^1$-coordinate, and $|\psi_0\rangle$ is the boundary state vector of the Hilbert space on $x^1$-coordinate with the spinor degrees of freedom.
The two component spinor $|s\rangle$ corresponding to edge bound states satisfies $\rmi\gamma^1|s\rangle={\rm sgn}(m)|s\rangle$, where ${\rm sgn}(m)$ indicates the sign of the mass in $x^1<0$, while 
the spinor satisfying $\rmi\gamma^1|s\rangle=-{\rm sgn}(m)|s\rangle$
corresponds to states that cannot be normalized.
By using the relation $\gamma^{\mu}\gamma^{\nu}=\eta^{\mu\nu}-i\epsilon^{\mu\nu\rho}\gamma_{\rho}$, the condition on $|s\rangle$ can be rewritten in a convenient form, $\gamma^0\gamma^2|s\rangle=-{\rm sgn}(m)|s\rangle$.

Next, we consider the Hamiltonian of the gapless boundary states resulting from the bulk Hamiltonian by projecting the Hilbert space on the $x^1$-coordinate onto that of the boundary mode. 
The projected Hamiltonian is given by defining the projection opeartor $P=|\psi_0\rangle\langle\psi_0|$ as $P^{\dagger}\mathcal{H}P$.
In the following, the $x^2$-dependent part $\tilde{\mathcal{H}}\equiv\langle\psi_0|P^{\dagger}\mathcal{H}P|\psi_0\rangle=\langle\psi_0|\mathcal{H}|\psi_0\rangle$ of the projected Hamiltonian is referred to as the boundary Hamiltonian.
The boundary Hamiltonian for the unperturbed bulk Hamiltonian (\ref{eq:Hamiltonian_unperturbed}) is
\begin{eqnarray}
 \tilde{\mathcal{H}}_0
 &\equiv
 \langle\psi_0| 
 \mathcal{H}_0 |
 \psi_0\rangle \nonumber\\
 &=
 \langle s|
 -\rmi v
 \gamma^0\gamma^2
 \partial_2
 |s\rangle \nonumber\\
 &=
 \rmi v\,
 {\rm sgn}(m)
 \partial_2.
 \label{eq:Hamiltonian_unperturbed_boundary}
\end{eqnarray}
Also we consider the projection of the perturbation term (\ref{eq:Hamiltonian_perturbation}) onto the boundary mode.
For convenience, the width of the boundary states in the $x^1$-direction is tuned to be narrow enough so that typical length scale of the gravitational field is much longer than the width of the boundary states.
This situation allows us to use
an assumption that the metric depends only on $x^2$ near the boundary.
Then $x^1$ and $x^2$ is completely decoupled in the boundary Hamiltonian.
The boundary Hamiltonian for the second term of the perturbation term (\ref{eq:Hamiltonian_perturbation}) is
\begin{eqnarray}
 \langle\psi_0|
 \rmi v (h^{\;\;j}_{\alpha}/2)
 \gamma^0\gamma^{\alpha}
 \partial_j
 |\psi_0\rangle 
 &=
 -
 \langle\psi_0|
 \gamma^0
 \gamma^{\alpha}
 \gamma^1
 m(x^1)
 |\psi_0\rangle  
 vh^{\;\;1}_{\alpha}(x^2)/2 \nonumber\\ 
 &\quad
 +
 \langle s|
 \gamma^0\gamma^{\alpha}
 |s\rangle
 \rmi v h^{\;\;2}_{\alpha}(x^2)\partial_2/2.
 \label{eq:Hamiltonian_perturbation_boundary}
\end{eqnarray}
The first term in (\ref{eq:Hamiltonian_perturbation_boundary}) is zero since it contains an integral
\begin{eqnarray}
 &\int_{-\infty}^{\infty} \rmd x^1
 m(x^1)
 \exp
 \left[
 2v^{-1}{\rm sgn}(m)
 \int_0^{x^1} \rmd{x'}^1\, m({x'}^1)
 \right]
 \nonumber\\
 &\propto
 \exp
 \left.
 \left[
 2v^{-1}{\rm sgn}(m)
 \int_0^{x^1} \rmd{x'}^1\, m({x'}^1)
 \right]
 \right|_{x^1=-\infty}^{x^1=\infty}=0.
\end{eqnarray}
The second term in (\ref{eq:Hamiltonian_perturbation_boundary}) is nonzero for $\alpha=0,2$ due to the property of the two-component spinor $|s\rangle$.
The boundary Hamiltonian for
the perturbation term is written as
\begin{eqnarray}
 \tilde{\mathcal{H}}_1
 \equiv
 \langle\psi_0| 
 \mathcal{H}_1
 |\psi_0\rangle
 =
 \zeta(x^2)
 (-\rmi v\partial_2),
 \label{eq:Hamiltonian_perturbation_boundary_2}
\end{eqnarray}
where $\zeta(x^2)=[-{\rm sgn}(m)h(x^2)+{\rm sgn}(m)h^{\;\;2}_{2}(x^2)-h^{\;\;2}_{0}(x^2)]/2$.
The boundary Hamiltonian for second perturbation term (\ref{eq:Hamiltonian_perturbation2})
\begin{eqnarray}
 \tilde{\mathcal{H}}_2
 &\equiv
 \langle\psi_0| 
 \mathcal{H}_2
 |\psi_0\rangle \nonumber\\
 &=
 -(\rmi v\,{\rm sgn}(m)/4)
 \left[
  \omega^{(1)}_{0\;20}
  +
  \omega^{(1)}_{1\;12}
 \right] 
 -
 (\rmi v/4)
 \left[
  \omega^{(1)}_{1\;01}
  -
  \omega^{(1)}_{2\;20}
 \right] \\
 &\equiv
 \chi(x^2). \nonumber
\end{eqnarray}

\subsection{Effective boundary free energy}
\label{sec:boundary_FE}

Here we calculate the effective free energy functional of the gravitational field of the boundary mode. 
Tracing out the boundary fermionic degrees of freedom, the second order perturbation term in (\ref{eq:effectiveFE_perturbation_2}) is given by 
\begin{eqnarray}
 -\beta F^{\rm{bdry}} 
 =
 \frac{1}{2}
 \sum_n
 \int \rmd x^2\,
 \langle x^2|
 \tilde{\mathcal{G}}_0
 \tilde{\mathcal{U}}
 \tilde{\mathcal{G}}_0
 \tilde{\mathcal{U}}
 |x^2\rangle,
\end{eqnarray}
where $\tilde{\mathcal{G}}_0=1/(-i\omega_n+\tilde{\mathcal{H}}_0)$ is the temperature Green's function of the boundary fermion, and 
$\tilde{\mathcal{U}}=\tilde{\mathcal{H}}_1+\tilde{\mathcal{H}}_2$.
Tracing over the Hilbert space on the $x^2$-coordinate, the effective free energy becomes
\begin{eqnarray}
 -\beta F^{\rm{bdry}} 
 &=
 \int\frac{\rmd q}{2\pi}
 \int \rmd x^2 \rmd{x'}^2
 \rme^{\rmi q(x^2-{x'}^2)} 
 \left[
  \Pi^{(2)}(q)
  \zeta(x^2)\zeta({x'}^2)\right. \nonumber\\
 &\qquad
 \left.
  +
  2\Pi^{(1)}(q)
  \zeta(x^2)\chi({x'}^2)
  +
  \Pi^{(0)}(q)
  \chi(x^2)\chi({x'}^2)
 \right],
 \label{eq:effectiveFE_2nd}
\end{eqnarray}
where $q$ is the momentum of the gravitational field, and $\Pi^{(s)}(q)$ is the one-loop integral of the single-component boundary fermion given by 
\begin{eqnarray}
 \Pi^{(s)}(q) 
 &=
 \frac{v^s}{2}
 \sum_n 
 \int \frac{\rmd p}{2\pi}
 \left(\frac{2p+q}{2}\right)^s \nonumber\\
 &\quad\times
 \frac{1}{-\rmi\omega_n+vp}
 \frac{1}{-\rmi\omega_n+v(p+q)}.
 \label{eq:effectiveFE_2nd_coefficient}
\end{eqnarray}
The leading contribution of the integral over $q$ in (\ref{eq:effectiveFE_2nd}) is of order $q^0$, where we expand the integral in the powers of $q$ since the long-wave length behavior of the gravitational field
 is of interest.
Summation over the Matsubara frequencies gives $\sum_n (-\rmi\omega_n+vp)^{-2}=\beta \rmd f(vp)/\rmd(vp)$, where $f(\epsilon)$ is the Fermi-Dirac distribution function at temperature $T$.
The coefficient (\ref{eq:effectiveFE_2nd_coefficient}) of the order of $q^0$ is given by
\begin{eqnarray}
 \Pi^{(s)}(0)
 &=
 \frac{\beta}{2v}
 \int \frac{\rmd p}{2\pi} 
 \frac{\rmd f(p)}{\rmd p} p^s.
\end{eqnarray}
Here we have replaced $vp$ by $p$ for simplicity.
Obviously $\Pi^{(1)}(0)=0$, since $\rmd f(p)/\rmd p$ is an even function of $p$. 
Furthermore, $\Pi^{(0)}(0)=-\beta/4\pi v$ can be neglected in this study since the free energy proportional to $\Pi^{(0)}(0)$ is independent of temperature when substituted into (\ref{eq:effectiveFE_2nd}), which indicates that this term is not relevant to the quantized thermal Hall effect.
The only coefficient that is concerned is
\begin{eqnarray}
 \Pi^{(2)}(0)
 =
 \frac{\beta}{v}
 \int \frac{\rmd p}{2\pi}
 \left[\theta(-p)-f(p)\right]p,
 \label{eq:effectiveFE_2nd_coefficient_2}
\end{eqnarray}
where $\theta$ is the Heaviside step function, and is regarded as the zero-temperature Fermi-Dirac distribution function. 
At low temperature, 
the distribution function in (\ref{eq:effectiveFE_2nd_coefficient_2}) can be expanded with respect to the temperature by the Sommerfeld expansion $f(p)\simeq\theta(-p)-(\pi^2T^2/6)(\rmd\delta(p)/\rmd p)$ as
\begin{eqnarray}
 \Pi^{(2)}(0)
 &\simeq
 \frac{\beta}{v}
 \int \frac{\rmd p}{2\pi}
 \frac{\pi^2T^2}{6}
 \frac{\rmd\delta(p)}{\rmd p}
 p 
 =
 -\beta
 \frac{\pi T^2}{12v}.
 \label{eq:effectiveFE_2nd_coefficient_3}
\end{eqnarray}
Then the effective free energy is reduced to a simple form as
\begin{eqnarray}
 F^{\rm{bdry}} 
 &=
 \frac{\pi T^2}{12v}
 \int_{-\infty}^{\infty} \rmd x^2 
 [\zeta(x^2)]^2 \nonumber\\
 &=
 \frac{\pi T^2}{48v}
 \int_{-\infty}^{\infty} \rmd x^2 
 [-{\rm sgn}(m)h+{\rm sgn}(m)h^2_{\;\;2}-h^2_{\;\;0}]^2. 
 \label{eq:effectiveFE_2nd_2}
\end{eqnarray}

\subsection{Bulk thermal Hall conductivity}
\label{sec:bulk_conductivity}

The energy current $j_E^{\rm{bdry}}$ flowing along the boundary can be read off from the boundary effective free energy (\ref{eq:effectiveFE_2nd_2}).
Using an equivalence between the energy current and the momentum density resulting from the Lorentz invariance,
the boundary energy current is $j_E^{\rm{bdry}}\equiv vT^{02}\simeq 2v (\delta F^{{\rm bdry}}/\delta h_{20})$, which results in
\begin{eqnarray}
 &j_E^{\rm{bdry}}
 =
 \frac{\pi T^2}{12}
 [-{\rm sgn}(m)h+{\rm sgn}(m)h^2_{\;\;2}-h^2_{\;\;0}].
 \label{eq:bulkBoundary_energyCurrent}
\end{eqnarray}

The Str\v{e}da formula for the quantized thermal Hall effect, the form of which is given in (\ref{eq:Streda_thermal_Hall}), implies that the thermal Hall conductivity is given by ratio of the variation of the energy magnetization and that of temperature.
Thus we consider a two-dimensional system with the boundary under a spatially \textit{homogeneous} gravitational potential $\phi$, and see how the bulk energy magnetization changes according to the change of the gravitational potential homogeneously.
For this purpose, only the first term of the right-hand side of (\ref{eq:bulkBoundary_energyCurrent}) is considered since this term is proportional to the gravitational potential $\phi$. 
The energy magnetization is defined by
\begin{eqnarray}
 \bm{j}_E=\bm{\partial}\times\bm{M}_E.
 \label{eq:energy_magnetization_definition}
\end{eqnarray}
Here, we assume the energy magnetization is uniformly distributed in the bulk as a response of the uniform bulk gravitational potential $\phi$. 
This assumption is same as the one used in the phenomenological theory, that is, the energy magnetization depends only on the temperature, which is noted just above (\ref{eq:Streda_thermal_Hall}).
Since uniform energy magnetization leads to vanishing bulk energy current due to (\ref{eq:energy_magnetization_definition}), the energy magnetization in the bulk can be evaluated entirely by the boundary energy current.

The bulk energy magnetization can be derived from the boundary energy current by the following argument. 
Consider a boundary at $x^1=0$ separating the bulk region with mass $m$ ($x^1<0$) and outside region with mass $-m$ ($x^1>0$).
The bulk energy magnetization at the region with mass $m$ and that with mass $-m$ have opposite sign since the energy magnetization changes its sign by parity transformation, which is same as the orbital magnetization in the quantum Hall system.
The boundary energy current is defined as the energy current flowing in $x^2$-direction in the boundary region where the energy magnetization changes from bulk value to one in the outer region. 
Then the energy magnetization $M^z$ in the bulk $x<0$ is given in terms of the boundary energy current as
\begin{eqnarray}
 j_E^{\rm{bdry}}
 &\equiv
 \int_{-\infty}^{\infty} \rmd x^1 j_E^2(x^1) \nonumber\\
 &=
 \int_{-\infty}^{\infty} \rmd x^1(-\partial_1 M_E^z(x^1)) \nonumber\\
 &=
 -\left(
 M_E^z(x^1=+\infty)-M_E^z(x^1=-\infty)
 \right) \nonumber\\
 &=
 2M^z_E,
\end{eqnarray}
which results in 
\begin{eqnarray}
 M^z_E
 =
 -{\rm sgn}(m)
 \frac{\pi T^2}{24}
 h.
 \label{eq:bulk_energy_magnetization}
\end{eqnarray}

Substituting $h=h^{\mu}_{\;\;\mu}=2\phi$ and using the Tolman-Ehrenfest relation (\ref{eq:Luttinger_temperature_gravity}), the thermal Hall conductivity is then given by the Str\v{e}da formula for the quantized thermal Hall effect:
\begin{eqnarray}
 \kappa_H
 =
 -
 \frac{1}{T}
 \frac{\partial M^z_E}{\partial \phi}
 =
 {\rm sgn}(m)
 \frac{\pi T}{12},
 \label{eq:thermal_Hall_conductivity}
\end{eqnarray}
which is the quantized thermal Hall conductivity for the Chern number $C={\rm sgn}(m)/2$.

\section{Bulk physical quantities}
\label{sec:bulk_properties}

In this section, we elaborate to make a connection between the boundary effective theory and the bulk quantities of the thermal Hall effect, that is, the thermal Hall current (\ref{eq:thermal_Hall_effect}) and the bulk effective free energy (\ref{eq:effective_phenomenology}).
The expression of the bulk thermal Hall current is deduced from the boundary energy current (\ref{eq:bulkBoundary_energyCurrent}), and we conjecture a possible form of the bulk effective free energy which creates a bulk quantized thermal Hall current.
It should be noted that while the bulk thermal Hall conductivity (\ref{eq:thermal_Hall_conductivity}) is a direct consequence of the boundary effective free energy (\ref{eq:effectiveFE_2nd_2}), results in this section may be regarded as somehow indirect consequence from the boundary theory, since it is unclear whether or not, at the thermal equilibrium, the thermal Hall current flows in the bulk or only along the boundary in a fully gapped two-dimensional insulator\cite{manes13,kane97,smrcka77}.

\subsection{Bulk thermal Hall current}
\label{sec:bulk_current}

The expression of the bulk energy current can be obtained by considering  a two-dimensional system with the boundary under spatially \textit{inhomogeneous} gravitational potential.
If the boundary is considered as an isolated one-dimensional system obeying the free energy (\ref{eq:effectiveFE_2nd_2}), the energy current (\ref{eq:bulkBoundary_energyCurrent}) can break the energy conservation law when the metric varies spatially: $\partial_2j_E^{\rm{bdry}}\neq 0$.
Here, note that the boundary energy current is not dependent on time, since the boundary free energy (\ref{eq:effectiveFE_2nd_2}) and responses derived from it are static.
The energy conservation law is retained by including incoming energy flow from the gapped bulk.
The bulk effective free energy is determined to precisely describe this compensating bulk energy current.
The incoming energy current from one side of the semi-infinite space of the bulk is equal to the other side, since the mass term is inverted by the time-reversal operation or the parity inversion, which causes inversion of the thermal Hall coefficient.
Thus,
the bulk energy current $j_E$ near the boundary satisfies the following equation:
\begin{eqnarray}
 j_E^1(x^1=-0)=-j_E^1(x^1=+0)=\frac{1}{2}\partial_2j_E^{\rm{bdry}}.
 \label{eq:energy_conservation}
\end{eqnarray}

In the following, we show that only the first term of the right-hand side of (\ref{eq:bulkBoundary_energyCurrent}) contributes to $j_E^{\rm{bdry}}$ in (\ref{eq:energy_conservation}) by taking into account that the bulk energy current, $j_E^1$ in (\ref{eq:energy_conservation}), should not be dependent on details of the boundary, since it results from the nontrivial topology of the bulk energy bands.
To see this, we examine how each term in (\ref{eq:effectiveFE_2nd_2}) changes according to the direction of the boundary.
The system is rotated by an angle $\theta$ anticlockwise to define a new boundary. 
New coordinate variables perpendicular and parallel to the boundary are introduced as
\begin{eqnarray}
 \left\{
 \begin{array}{l}
  {x'}^1=x^1\cos\theta+x^2\sin\theta \\
  {x'}^2=-x^1\sin\theta+x^2\cos\theta
 \end{array}
 \right.
 \,\,\Leftrightarrow\,\,
 {x'}^j=R^j_{\;\;k}(\theta)x^k,
\end{eqnarray}
where $R^j_{\;\;k}(\theta)$ is the rotation matrix.
Derivatives in the new coordinates are given by
\begin{eqnarray}
 \partial'_j
 =
 R_j^{\;\;k}(\theta)
 \partial_k,
\end{eqnarray}
where $R_j^{\;\;k}(\theta)$ is the inverse of $R^j_{\;\;k}(\theta)$ ($R_j^{\;\;k}(\theta)=R^k_{\;\;j}(-\theta)$, and $R_j^{\;\;k}(\theta)R^j_{\;\;l}(\theta)=\delta_l^k$ holds).
The Hamiltonian (\ref{eq:Hamiltonian_unperturbed}) and (\ref{eq:Hamiltonian_perturbation}) has SO(2) rotational symmetry, that is, introducing new vectors and tensors by
\begin{eqnarray}
 {\gamma'}^j
 =
 R^j_{\;\;k}(\theta)\gamma^k,
\end{eqnarray}
and
\begin{eqnarray}
 &{h'}^j_{\;\;0}(x)
 =
 R^j_{\;\;l}(\theta)h^l_{\;\;0}(x),
 \label{eq:h_temporal_rotation}\\
 &{h'}^j_{\;\;k}(x)
 =
 R^j_{\;\;l}(\theta)R_{k}^{\;\;m}(\theta)h^l_{\;\;m}(x),
 \label{eq:h_spatial_rotation}
\end{eqnarray}
the Hamiltonian is a scalar quantity with respect to the rotation $R^j_{\;\;k}(\theta)$: $\mathcal{H}_0[\gamma',h';x']=\mathcal{H}_0[\gamma,h;x]$ and $\mathcal{U}[\gamma',h';x']=\mathcal{U}[\gamma,h;x]$.
Note that $\gamma^0$ and other scalar quantities contained in the Hamiltonian are unchanged during the rotation. 
Then the effective free energy of the new boundary perpendicular to ${x'}^1$ is given by
\begin{eqnarray}
 &{F^{\rm{bdry}}}'
 =
 \frac{\pi T^2}{12v}
 \int_{-\infty}^{\infty} \rmd{x'}^2 [\zeta'({x'}^2)]^2.
 \label{eq:effectiveFE_2nd_3}
\end{eqnarray}
Here, $\zeta'({x'}^2)=[-{\rm sgn}(m)h({x'}^2)+{\rm sgn}(m){h'}^2_{\;\;2}({x'}^2)-{h'}^2_{\;\;0}({x'}^2)]/2$, and thus the boundary energy current flowing along the ${x'}^2$-coordinate is
\begin{eqnarray}
 {j_E^{\rm{bdry}}}'
 &=
 \frac{\pi T^2}{12}
 [-{\rm sgn}(m)h+{\rm sgn}(m){h'}^2_{\;\;2}-{h'}^2_{\;\;0}].
 \label{eq:boundary_energy_current_2}
\end{eqnarray}

The second and third terms of the right-hand side of (\ref{eq:boundary_energy_current_2}) explicitly depend on the boundary angle $\theta$ through (\ref{eq:h_temporal_rotation}) and (\ref{eq:h_spatial_rotation}), while the first term does not.
The bulk energy current, which is proportional to the Chern number of the bulk energy band, should not be dependent on the details of the boundary, such as the boundary angle.
Therefore the first term of the right-hand side of (\ref{eq:boundary_energy_current_2}) can be regarded as a term related to the bulk energy current by (\ref{eq:energy_conservation}).
The same conclusion can be given by considering the number of indices in (\ref{eq:energy_conservation}).
The right-hand side of the (\ref{eq:energy_conservation}) must be a vector quantity with respect to rotation since the left-hand side is a vector quantity.
This condition is satisfied when $j_E^{{\rm bdry}}$ is a scalar quantity,
which results in the first term of the right-hand side of (\ref{eq:boundary_energy_current_2}).
The bulk energy current corresponding to the boundary energy current $j_E^{{\rm bdry}}=-{\rm sgn}(m)(\pi T^2/12)h$ is given by
\begin{eqnarray}
 j_E^k
 =
 -{\rm sgn}(m)\frac{\pi T^2}{24}\epsilon^{kl}\partial_l h,
 \label{eq:bulk_energy_current}
\end{eqnarray}
the form of which is independent of the direction of  the boundary.
The equation (\ref{eq:bulk_energy_current}) exactly represents the quantized thermal Hall effect (\ref{eq:thermal_Hall_effect}) by substituting $h=2\phi$ from the metric (\ref{eq:metric_potential_rotation}), identifying the energy current with the thermal current, and using the Tolman-Ehrenfest relation (\ref{eq:Luttinger_temperature_gravity}).
The thermal Hall coefficient is given by
\begin{eqnarray}
 \kappa_H
 =
 -\frac{1}{T}
 \frac{j^{x}_E}{\partial_y \phi}
 =
 {\rm sgn}(m)\frac{\pi T}{12},
\end{eqnarray}
which is quantized by the Chern number $C={\rm sgn}(m)/2$.

\subsection{Effective bulk free energy}
\label{sec:bulk_FE}

The bulk free energy for the quantized thermal Hall effect can be deduced so that it realizes the energy current (\ref{eq:bulk_energy_current}) as a gravitational response, and is given by
\begin{eqnarray}
 F[h]
 &=
 -{\rm sgn}(m)\frac{\pi T^2}{48v}
 \int \rmd^2x\,
 \epsilon^{kl}
 h_{k0}
 \partial_l
 h.
 \label{eq:effectiveFE_2nd_5}
\end{eqnarray}
A similar form of the bulk gravitational free energy has also been predicted in Ref.~\cite{shitade14}. The free energy (\ref{eq:effectiveFE_2nd_5}) is regarded as a static and Lorentz invariant version of one in Ref.~\cite{shitade14}.

Total energy conservation can be seen by the translation symmetry of the bulk free energy together with the boundary free energy in the temporal direction. 
Under an infinitesimal coordinate transformation 
\begin{eqnarray}
 {x'}^0\to x^0+\zeta^0 \label{eq:translation_temporal},
\end{eqnarray} 
the metric varies as $\delta h_{k0}=\partial_k\zeta_0$, 
where we assume $\zeta^0$ is a function of $x^1$ and $x^2$ since we consider a static theory in this study.
The translation (\ref{eq:translation_temporal}) leaves the bulk free energy (\ref{eq:effectiveFE_2nd_5}) on a region $D$ unchanged except the boundary term on $\partial D$ given by
\begin{eqnarray}
 \delta F[h]
 &=
 -{\rm sgn}(m)\frac{\pi T^2}{48v}
 \int_D \rmd^2x\,
 \epsilon^{kl}
 (\partial_k\zeta_0)
 (\partial_l h) \nonumber\\
 &=
 -{\rm sgn}(m)\frac{\pi T^2}{48v}
 \int_{\partial D}\rmd\bm{l}\cdot
 (\zeta_0\bm{\partial} h).
 \label{eq:bulk_var}
\end{eqnarray}
On the other hand, by defining the boundary parallel to $x^2$ and the bulk in $x^1<0$, the translation (\ref{eq:translation_temporal}) makes a change in the boundary free energy as
\begin{eqnarray}
 \delta
 \left(
 \tilde{F}^{\rm{bdry}}/2
 \right)
 &=
 \delta 
 \left(
 -{\rm sgn}(m)\frac{\pi T^2}{48v}
 \int \rmd x^2\,h h_{20}
 \right) \nonumber\\
 &=
 {\rm sgn}(m)\frac{\pi T^2}{48v}
 \int \rmd x^2\,\zeta_0 \partial_2 h,
 \label{eq:boundary_var}
\end{eqnarray}
which precisely cancels the boundary term in (\ref{eq:bulk_var}).
Therefore the total free energy is invariant under the translation (\ref{eq:translation_temporal}), and it indicates the energy of the bulk and the boundary is totally conserved.
The boundary free energy $\tilde{F}^{\rm{bdry}}$ considered here is a part of (\ref{eq:effectiveFE_2nd_2}) that represents the scalar boundary energy current shown above (\ref{eq:bulk_energy_current}), while the other terms are irrelevant to the variation (\ref{eq:translation_temporal}) ($h^2$, $hh^2_{\;\;2}$, and $(h^2_{\;\;2})^2$) or to the bulk thermal Hall current ($(h^2_{\;\;0})^2$ and $h^2_{\;\;2}h^2_{\;\;0}$).
Furthermore the boundary free energy in the parenthesis of the left-hand side of (\ref{eq:boundary_var}) is a half of it, since the other half corresponds to the other side of the bulk ($x^1>0$).

In the presence of a gravitational potential and a rotation (\ref{eq:metric_potential_rotation}), $h$ and $h_{k0}$ in (\ref{eq:effectiveFE_2nd_5}) are given by $h=2\phi$ and $h_{k0}=(\Omega x^2/v, -\Omega x^1/v)$.
Therefore the integrand of (\ref{eq:effectiveFE_2nd_5}) together with that of half of the boundary free energy is $\epsilon^{kl}h\partial_kh_{l0}=-4\phi\Omega/v$.
The effective free energy under the metric (\ref{eq:metric_potential_rotation}) is given by 
\begin{eqnarray}
 F[\phi,\Omega]
 +
 \tilde{F}^{\rm{bdry}}[\phi,\Omega]/2 
 &=
 -{\rm sgn}(m)\frac{\pi T^2}{48v}
 \int \rmd^2x\,
 \epsilon^{kl}h\partial_kh_{l0} \nonumber\\
 &=
 \frac{\kappa_H T}{v^2}
 \int \rmd^2x\,
 \phi\Omega.
 \label{eq:effectiveFE_2nd_6}
\end{eqnarray}
The free energy (\ref{eq:effectiveFE_2nd_6}) realizes the coupling of the gravitational potential and the angular momentum, which is the same form as the phenomenologically derived free energy presented in (\ref{eq:effective_phenomenology}).
However, it turns out that, from the field theoretical method, the coefficient in (\ref{eq:effectiveFE_2nd_6}) is half of that predicted in \cite{nomura12}.

\section{Discussion}
\label{sec:discussion}

Using the finite-temperature field theoretical method, we have derived microscopically a boundary effective free energy functional of the gravitational field, and show that this free energy accounts for the quantized thermal Hall effect of a gapped fermionic system in two-dimension.
This result gives a novel insight into the effective theory of the quantized thermal Hall effect that the \textit{boundary} effective theory, derived microscopically from the gapped fermionic theory, is sufficient to account for the bulk quantized thermal Hall \textit{conductivity}, unlike previous studies of bulk effective theories for the quantized thermal Hall effect and the case of the quantum Hall effect.
However, connecting the boundary effective theory with the expression of the bulk thermal Hall \textit{current} and the bulk effective theory requires further speculation. 
After making this connection, our boundary effective theory is shown to be consistent with some of previous bulk effective theories\cite{nomura12,shitade14}.

The method we have used to study the effective theory is valid for a system in thermal equilibrium.
Thus the thermal current and transport coefficient derived from our effective theory are considered as those for equilibrium responses.
Obviously, chiral boundary energy current in a topological insulator can flow even in thermal equilibrium as the boundary electric current does. 
Thus there is no doubt that the quantized thermal Hall coefficient in the bulk can be derived from our boundary theory, since it depends only on the boundary energy current.
On the other hand, it is not yet clear whether the bulk thermal Hall current flows in thermal equilibrium or not. 
In our study, however, we have also given the expression of the bulk thermal Hall current within the equilibrium thermal response.

\section{Conclusion}
\label{sec:conclusion}

In this study, we have considered the Hamiltonian of the two-dimensional massive Dirac fermion under a static gravitational field which contains the gravitational potential mimicking temperature gradient, and the angular velocity coupling to the momentum density.
We place this model in a semi-infinite plane with boundary, and consider thermal equilibrium at a finite temperature.

At first, the bulk Hamiltonian is projected onto the boundary.
Then the boundary free energy is calculated from the chiral gapless boundary fermion by tracing out the fermionic degrees of freedom and perturbatively expanding the free energy with respect to the metric.
We have shown that the gravitational response of the boundary free energy at the boundary determines the bulk energy magnetization, and it explains the quantized thermal Hall conductivity through the Str\v{e}da formula.

Furthermore, the bulk energy current is evaluated from the boundary energy current to ensure that the total energy is conserved in the whole system. 
The resulting bulk energy current precisely represents the thermal Hall effect.
The effective free energy functional (\ref{eq:effectiveFE_2nd_5}) for the two-dimensional massive Dirac fermion is then deduced to realize the bulk energy current.
The bulk effective free energy reproduces the same form as the phenomenologically derived effective free energy (\ref{eq:effective_phenomenology}) by substituting a metric of a gravitational potential in a rotating frame. 

\ack
This work was supported by World Premier International Research Center
Initiative (WPI) and Grant-in-Aid for Scientific
Research (No.~25103703, No.~26107505 and No.~26400308) from MEXT,
the NSF under Grant No.~DMR-1455296, and Alfred P.~Sloan foundation.


\begin{thebibliography}{10}

\bibitem{hasan10}
M.~Z. Hasan and C.~L. Kane.
\newblock Colloquium: Topological insulators.
\newblock {\em Rev. Mod. Phys.}, 82:3045, 2010.

\bibitem{qi11}
Xiao-Liang Qi and Shou-Cheng Zhang.
\newblock Topological insulators and superconductors.
\newblock {\em Rev. Mod. Phys.}, 83:1057, 2011.

\bibitem{prange87}
R.~E. Prange and S.~M. Girvin, editors.
\newblock {\em The Quantum Hall Effect}.
\newblock Springer, 1987.

\bibitem{thouless82}
D.~J. Thouless, M.~Kohmoto, M.~P. Nightingale, and M.~den Nijs.
\newblock Quantized hall conductance in a two-dimensional periodic potential.
\newblock {\em Phys. Rev. Lett.}, 49:405, 1982.

\bibitem{read00}
N.~Read and Dmitry Green.
\newblock Paired states of fermions in two dimensions with breaking of parity
  and time-reversal symmetries and the fractional quantum hall effect.
\newblock {\em Phys. Rev. B}, 61:10267, 2000.

\bibitem{schnyder08}
Andreas~P. Schnyder, Shinsei Ryu, Akira Furusaki, and Andreas W.~W. Ludwig.
\newblock Classification of topological insulators and superconductors in three
  spatial dimensions.
\newblock {\em Phys. Rev. B}, 78:195125, 2008.

\bibitem{kitaev09}
Alexei Kitaev.
\newblock Periodic table for topological insulators and superconductors.
\newblock {\em AIP Conf. Proc.}, 1134:22, 2009.

\bibitem{ryu10}
Shinsei Ryu, Andreas~P Schnyder, Akira Furusaki, and Andreas W~W Ludwig.
\newblock Topological insulators and superconductors: tenfold way and
  dimensional hierarchy.
\newblock {\em New J. Phys.}, 12:065010, 2010.

\bibitem{bertlmann96}
R.~Bertlmann.
\newblock {\em Anomalies in Quantum Field Theory}.
\newblock Oxford University Press, Oxford, 1996.

\bibitem{fujikawa04}
K.~Fujikawa and H.~Suzuki.
\newblock {\em Path Integrals and Quantum Anomalies}.
\newblock Oxford University Press, Oxford, 2004.

\bibitem{qi08}
Xiao-Liang Qi, Taylor~L. Hughes, and Shou-Cheng Zhang.
\newblock Topological field theory of time-reversal invariant insulators.
\newblock {\em Phys. Rev. B}, 78:195424, 2008.

\bibitem{wang11}
Zhong Wang, Xiao-Liang Qi, and Shou-Cheng Zhang.
\newblock Topological field theory and thermal responses of interacting
  topological superconductors.
\newblock {\em Phys. Rev. B}, 84:014527, 2011.

\bibitem{ryu12}
Shinsei Ryu, Joel~E. Moore, and Andreas W.~W. Ludwig.
\newblock Electromagnetic and gravitational responses and anomalies in
  topological insulators and superconductors.
\newblock {\em Phys. Rev. B}, 85:045104, 2012.

\bibitem{hughes11}
Taylor~L. Hughes, Robert~G. Leigh, and Eduardo Fradkin.
\newblock Torsional response and dissipationless viscosity in topological
  insulators.
\newblock {\em Phys. Rev. Lett.}, 107:075502, 2011.

\bibitem{gromov15}
Andrey Gromov and Alexander~G. Abanov.
\newblock Thermal hall effect and geometry with torsion.
\newblock {\em Phys. Rev. Lett.}, 114:016802, 2015.

\bibitem{bradlyn15}
Barry Bradlyn and N.~Read.
\newblock Low-energy effective theory in the bulk for transport in a
  topological phase.
\newblock {\em Phys. Rev. B}, 91:125303, 2015.

\bibitem{alvarez-gaume84}
Luis Alvarez-Gaum\'e and Edward Witten.
\newblock Gravitational anomalies.
\newblock {\em Nucl. Phys. B}, 234:269, 1984.

\bibitem{vuorio86}
I.~Vuorio.
\newblock Parity violation and the effective gravitational action in three
  dimensions.
\newblock {\em Phys. Lett. B}, 175:176, 1986.

\bibitem{goni86}
M.~A. Goni and M.~A. Valle.
\newblock Massless fermions and (2+1)-dimensional gravitational effective
  action.
\newblock {\em Phys. Rev. D}, 34:648, 1986.

\bibitem{ojima89}
Shuichi Ojima.
\newblock Derivation of gauge and gravitational induced chern-simons terms in
  three dimensions.
\newblock {\em Prog. Theor. Phys.}, 81:512, 1989.

\bibitem{stone12}
Michael Stone.
\newblock Gravitational anomalies and thermal hall effect in topological
  insulators.
\newblock {\em Phys. Rev. B}, 85:184503, 2012.

\bibitem{luttinger64}
J.~M. Luttinger.
\newblock Theory of thermal transport coefficients.
\newblock {\em Phys. Rev.}, 135:A1505, 1964.

\bibitem{qin11}
Tao Qin, Qian Niu, and Junren Shi.
\newblock Energy magnetization and the thermal hall effect.
\newblock {\em Phys. Rev. Lett.}, 107:236601, 2011.

\bibitem{nomura12}
Kentaro Nomura, Shinsei Ryu, Akira Furusaki, and Naoto Nagaosa.
\newblock Cross-correlated responses of topological superconductors and
  superfluids.
\newblock {\em Phys. Rev. Lett.}, 108:026802, 2012.

\bibitem{sumiyoshi13}
Hiroaki Sumiyoshi and Satoshi Fujimoto.
\newblock Quantum thermal hall effect in a time-reversal-symmetry-broken
  topological superconductor in two dimensions: Approach from bulk
  calculations.
\newblock {\em J. Phys. Soc. Jpn.}, 82:023602, 2013.

\bibitem{streda82}
P~Streda.
\newblock Theory of quantised hall conductivity in two dimensions.
\newblock {\em J. Phys. C: Solid State Phys.}, 15:L717, 1982.

\bibitem{landau75}
L.~D. Landau and E.~M. Lifshitz.
\newblock {\em The Classical Theory of Fields}.
\newblock Butterworth-Heinemann, 1975.

\bibitem{niemi83}
A.~J. Niemi and G.~W. Semenoff.
\newblock Axial-anomaly-induced fermion fractionization and effective
  gauge-theory actions in odd-dimensional space-times.
\newblock {\em Phys. Rev. Lett.}, 51:2077, 1983.

\bibitem{redlich84l}
A.~N. Redlich.
\newblock Gauge noninvariance and parity nonconservation of three-dimensional
  fermions.
\newblock {\em Phys. Rev. Lett.}, 52:18, 1984.

\bibitem{redlich84d}
A.~N. Redlich.
\newblock Parity violation and gauge noninvariance of the effective gauge field
  action in three dimensions.
\newblock {\em Phys. Rev. D}, 29:2366, 1984.

\bibitem{babu87}
K.~S. Babu, Ashok Das, and Prasanta Panigrahi.
\newblock Derivative expansion and the induced chern-simons term at finite
  temperature in 2+1 dimensions.
\newblock {\em Phys. Rev. D}, 36:3725, 1987.

\bibitem{manes13}
Juan~L. Ma{\~n}es and Manuel Valle.
\newblock Parity odd equilibrium partition function in 2 + 1 dimensions.
\newblock {\em J. High Energy Phys.}, 11:178, 2013.

\bibitem{kane97}
C.~L. Kane and Matthew P.~A. Fisher.
\newblock Quantized thermal transport in the fractional quantum hall effect.
\newblock {\em Phys. Rev. B}, 55:15832, 1997.

\bibitem{smrcka77}
L~Smr\v{c}ka and P~St\v{r}eda.
\newblock Transport coefficients in strong magnetic fields.
\newblock {\em J. Phys. C: Solid State Phys.}, 10:2153, 1977.

\bibitem{shitade14}
Atsuo Shitade.
\newblock Heat transport as torsional responses and keldysh formalism in a
  curved spacetime.
\newblock {\em Prog. Theor. Exp. Phys.}, page 123I01, 2014.

\end{thebibliography}
\end{document}